\documentclass[aps,amsmath,preprint]{revtex4}

\usepackage{graphicx}
\usepackage{subfigure}

\begin{document}

\title{Enhancement of two photon processes in quantum dots embedded in subwavelength
metallic gratings}
\author{Moshe G. Harats}
\author{Ronen Rapaport}
\affiliation{Racah Institute of Physics, The Hebrew University of Jerusalem, Jerusalem 91904, Israel}
\author{Adiel Zimran}
\author{Uri Banin}
\affiliation{Institute of Chemistry and The Center for Nanoscience and Nanothechnology, The Hebrew University of Jerusalem, Jerusalem 91904, Israel}
\author{Gang Chen}
\affiliation{Bel Laboratories, Alcatel Lucent, 600 Mountain Ave. Murray Hill, NJ 07974}
\begin{abstract}

We show a large enhancement of two-photon absorption processes in nanocrystal quantum dots and of light upconversion efficiency from the IR to the near-IR spectral regime, using a hybrid optical device in which near-IR emitting InAs quantum dots were embedded on top a metallic nanoslit array. The resonant enhancement of these nonlinear optical processes is due to the strong local electromagnetic field enhancements inside the nanoslit array structure at the extraordinary transmission resonances. A maximal two-photon absorption enhancement of more than 20 was inferred. Different high field regions were identified for different polarizations, which can be used for designing and optimizing efficient nonlinear processes in such hybrid structures. Combining nanocrystal quantum dots with subwavelength metallic nanostructures is therfore a promising way for a range of possible nonlinear optical devices.
\end{abstract}

\maketitle

\section{Introduction}

Nanocrystal Quantum Dots (NQDs) are becoming increasingly important as the active building blocks for a host of photonic applications. This is due to their unique properties, such as their well engineered optical emission and absorption wavelengths, their high quantum yield and their tiny dimensions. These properties, together with the ability to selectively integrate them on various platforms down to the nanoscale by changing their chemical functionality, make them a suitable active medium for many nano-optical devices.
Indeed,  NQDs are currently being used or tested for applications ranging from biological
labeling \textit{in vivo }of cells \cite{TPA-micro1,TPA-micro2,TPA-micro3,TPA-micro4},
photodetectors \cite{QD_photodetector}, solar cells \cite{QD_solar1,QD_solar2},
to lasing devices \cite{QD_lasing1,QD_lasing2,QD_lasing3} as well as for single photon
sources \cite{single_photon1,single_photon2}.

For improving and expanding the range of possible optical functionalities of NQDs in optical devices, control and manipulation schemes of their optical properties are necessary. Indeed, a demonstration of control of quantum dots linear emission and absorption has been demonstrated in various works \cite{qd_enhance_proton,qd_enhance_microcavity,qd_enhance_SPP,oron}. Some of the most interesting possible applications of NQDs in optical devices involve exploiting their optical nonlinear properties \cite{tpa_size_experiment, tpa_size_theory, TPA_size_strong_confinement}. For efficient nonlinear optical processes involving NQDs, an enhancement of their nonlinear response is essential. In this regard, several works have demonstrated enhancement of two-photon absorption (TPA) processes of quantum dots using local field enhancements in dielectric photonic crystals \cite{qd_TPA_PC_exp,QD_TPA_PC_theor}. Such ability to combine NQDs on the ensemble level as well as on the single particle level into photonic structures, and utilize it for manipulating the light - quantum dot interactions, opens up ways to construct various types of active optical devices, both linear and nonlinear, in subwavelengths dimensions and with new or enhanced properties.

 Another emerging set of optical structures for manipulation of light are subwavelength metallic structures \cite{Garcia_vidal_prediction,Garcia_vidal_review}. More specifically, metallic nanoslit array (NSA) structures are composed of metal transmission gratings with very narrow slits. Such structures have been predicted \cite{Garcia_vidal_prediction} and later shown \cite{first_grating_experiment} to exhibit a phenomenon known as Extra Ordinary Transmission (EOT) \cite{Ebbesen1998}, in which, under certain resonant wavelength regimes, light can be transmitted in an almost perfect way through these gratings. This almost perfect transmission occurs even when the slits are much smaller than the impinging light wavelength. An intuitive and clear explanation of the exact mechanism behind this effect was, and still is in some way, a matter of an ongoing scientific exploration, but it is now well established that these resonances correspond to standing surface waves on the NSA \cite{Garcia_vidal_review}. Initially, only resonances for incoming light in TM polarization were suggested and measured in relation to the EOT effect, but later, EOT was suggested \cite{garcia_vidal_TE_predict} and measured \cite{Garci-vidal2-s-polarization} in NSA structures for incoming light with TE polarization as well. For an EOT to occur in TE, a thin layer of dielectric material has to be present on top of the metallic grating (see a schematic diagram of such a structure in Fig. \ref{fig:exper_setup}a).

 Recently, we have developed a unified analytical model to explain in an intuitive way the underlying mechanism behind EOT in such structures, for both the TM and TE polarization on the same footing \cite{Ilai}. The emerging picture from this model is that in order for an EOT resonance to occur, the bragg waves, that are the confined eigenmodes of the periodic NSA, must fulfill a standing wave condition between the effective edges of the structure. Whenever such a condition is fulfilled, a standing waveguide mode which can also be viewed as a Fabri-Perot - like mode arises in the structure, \textit{leading to strong local electromagnetic field enhancements} at the anti-nodes of this standing wave \cite{Garcia_vidal_review,shen_platzman}. Local field enhancements were recently used to show a transmission bistability for a nonlinear polymer on top of an array of nanosized holes \cite{bistability_exp}. This picture is true for both TM and TE polarization with the difference being in the boundaries of the standing wave \cite{Ilai}, and therefore in the location of the EM field enhancements as will be discussed late on in this paper.

TPA is a third order nonlinear effect ($\chi^{(3)}$) which is an intensity dependent process. Therefore, it is expected that enhancing the electromagnetic (EM) field intensities at the vicinity of the nonlinear material will increase the TPA efficiency. Indeed, increasing the efficiency of two-photon absorption in Erbium doped glass using an NSA structure in TM polarization has recently been reported \cite{up_conversion}. In this work we show an enhancement of TPA and of light upconversion in NQDs embedded in a metallic subwavelength nanoslit array.
We show that the highly confined EM waves \textit{in both TE and TM polarizations} in such metallic structures enhances TPA efficiency in NQDs and therefore the light upconversion process by a factor of $\sim20$. We also show that the TPA enhancement measurement is a good probe of the local field enhancements in such metallic nanostructures, and suggest that understanding and exploiting the different geometry of the EOT resonances in TE and TM can lead to new functionalities of devices based on NQDs.

\section{Experimental setup}

\begin{figure}
\centering{}\includegraphics[width=0.8\columnwidth]{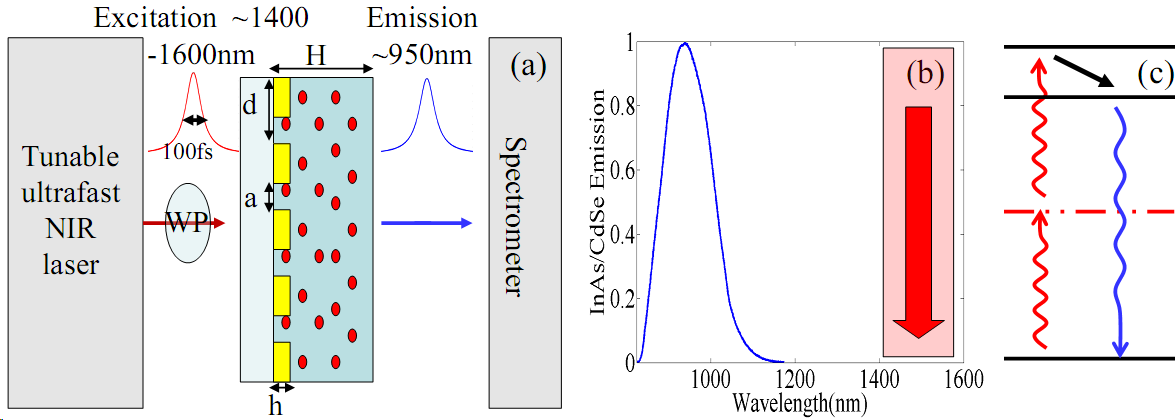}\caption{\label{fig:exper_setup}(a) A schematic of the experimental setup: the NSA structure depicted in the center consists of a dielectric polymer layer with NQDs (marked by red circles) on top of an Al grating. The red pulse refers to the exciting NIR pulse while the blue pulse refers
to the upconversion light emitted from the NQDs. The NSA sample dimensions are given by: a - slit width, d - NSA periodicity, h - Aluminum height, H - width of NQDs-in-polymer layer. (b) The emission spectra of the NQDs (blue line) and the spectral range of the laser excitation wavelengths (red box). (c) An energy level scheme of TPA and the resulting upconversion process.}

\end{figure}

To demonstrate the enhancement of a TPA process in NQDs using resonant local field enhancements in metallic NSAs, we fabricated metallic NSA structures covered by a NQD-in-polymer matrix. The NSA resonances were designed to occur at wavelengths which are just above half the first excitonic transition of the NQDs, tuned at half the wavelength of the higher excitonic transitions of the NQDs (see Fig. \ref{fig:exper_setup}c). By exciting the structure with laser pulses at these resonant wavelengths ($\sim 1500nm$), the strong local field enhancements of the structure enhanced the TPA process in the NQDs, leading to a population of the NQD excited states. These excited states then decay radiatively from the first excitonic transition wavelength ($\sim 950nm$), leading to light upconversion from the structure. This procedure is schematically illustrated in \ref{fig:exper_setup}c.

Our samples consist of NIR emitting NQDs in a transparent polymer, embedded in an Aluminum based NSA, as is schematically shown in Fig. \ref{fig:exper_setup}a. The NQDs investigated in this work are InAs/CdSe core/shell type I \cite{Banin_fab}, dispersed in a polymer matrix of perfluorocyclobutane (PFCB) \cite{Ronen1}. A typical emission spectrum of the NQDs is shown in Fig \ref{fig:exper_setup}b, peaking around $\sim 950nm$, corresponding to the emission wavelength of the first excitonic transition, inhomogeneously broadened by the small size distribution of the NQDs. The polymer has a refractive index of $n\simeq1.51$. The NQD/polymer mixture was
spin-coated on the NSA forming a $H\simeq1.8\mu m$
thick layer (Fig. \ref{fig:exper_setup}b).

The NSA consists of an Aluminum grating on a glass substrate, as depicted in Fig. \ref{fig:exper_setup}a with a height of $h=0.25\mu m$. Two different grating periods and slit widths were used, depending on the incoming excitation polarization with respect to the grating (either TE or TM polarizations). For the experiments with an excitation having TM (TE) polarization, the grating had a period of $d=0.8\mu m$ ($d=1\mu m$) and a slit width of $a=0.35\mu m$ ($a=0.45\mu m$) respectively. These parameters where chosen so that the structure resonances (in both TE and TM) would match the excitation wavelengths required for the TPA process as described above.

The excitation was done using a tunable Optical Parametric Oscillator
(OPO) pumped by a mode-locked 80Mhz Ti:Sapphire laser, producing sub-100fs pulses in the near infra-red (NIR) spectral range ($1400-1600nm$). The pulse
was either TM or TE polarized with respect to the NSA structure. The pulse was slightly focused on the sample, and the resulting upconverted fluorescence was collected to and analyzed by a spectrometer with a Pixis CCD camera (Fig. \ref{fig:exper_setup}a). This fluorescence intensity was normalized to the fluorescence intensity from an identical reference sample but without the NSA.

\section{Results}

In order to analyze our experimental results, We first define the total absorption
cross-section of the NQDs as: \begin{equation}
\sigma(\lambda,I)=\sigma_{0}(\lambda)+\sigma^{(2)}(\lambda) I\label{eq:total_abs}\end{equation}
where $\sigma_{0}(\lambda)$ is due to the linear resonant absorption of the NQD optical transitions, and $\sigma^{(2)}(\lambda)$
is the cross-section of the TPA of the incoming light at half the NQD excitonic optical transition (with intensity $I$).
As under such excitation conditions $\sigma_{0}$ vanishes, the total cross-section depends linearly on the excitation intensity.
The upconverted fluorescence intensity $I_{UC}$ is proportional to the number
of NQDs excited by TPA, defined as $N_{NQD}=\frac{\sigma I}{\hbar\omega}$ yielding \cite{Boyd}:\begin{equation}
I_{UC}\propto N_{NQD}=\frac{\sigma^{(2)}I^{2}}{\hbar\omega}\label{eq:quadratic}\end{equation}

Eq. \ref{eq:quadratic} shows that $I_{UC}\propto I^{2}$. We also define the enhancement factor of the incoming light inside the NSA structure for various excitation wavelengths $\lambda$ as $\gamma(\lambda)$. Therefore, the intensity inside the structure is enhanced as $I\rightarrow\gamma I$. As a result, the measured upconverted fluorescence  $I_{UC}$ from the NSA structure, normalized to the fluorescence from the reference without the NSA, is equal to $\gamma^2$.

\begin{figure}
\begin{centering}
\includegraphics[width=0.4\columnwidth]{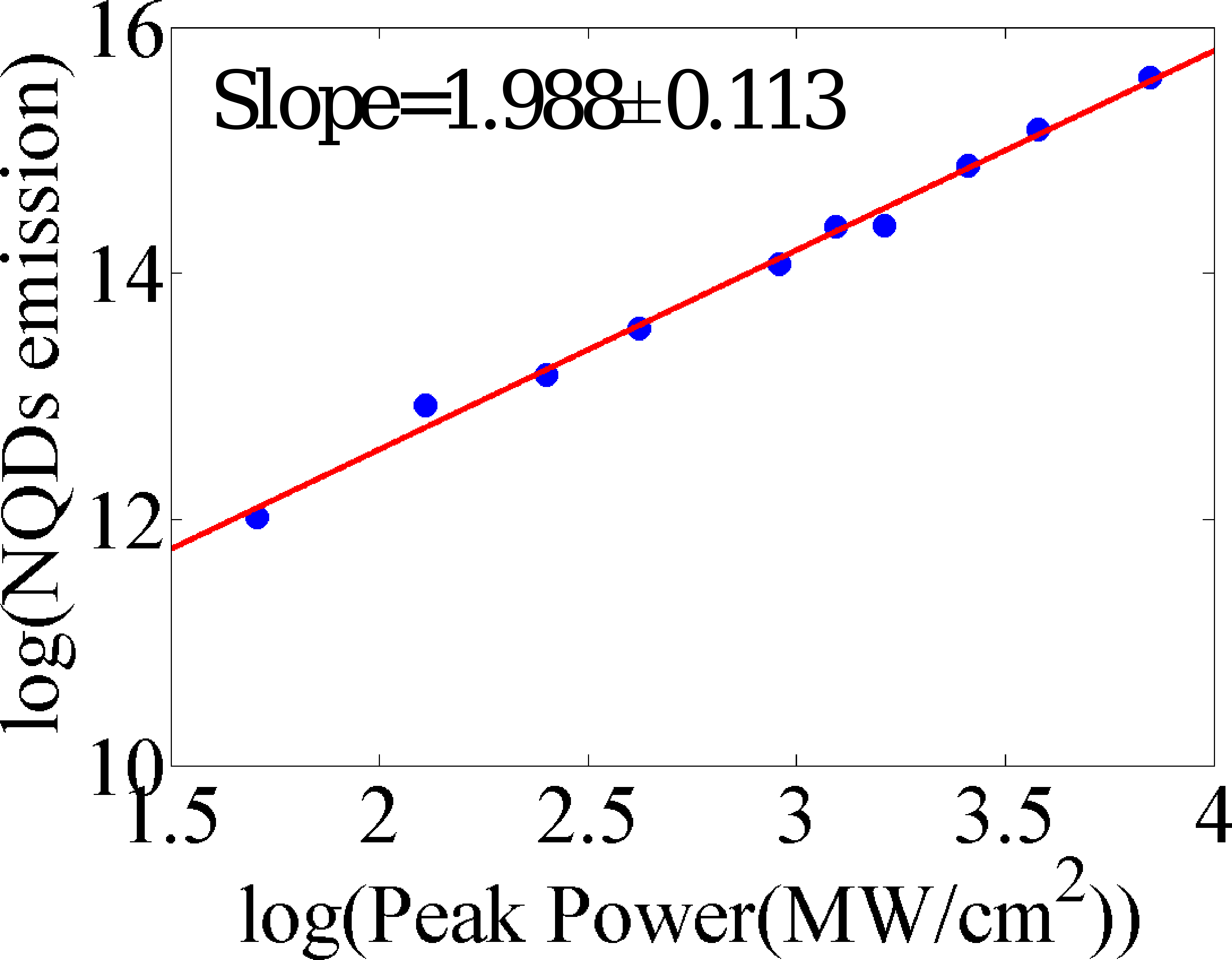}\caption{\label{fig:Quadratic}A logarithmic scale of $I_{UC}$ at an excitation wavelength
of $1505nm$. The upconversion due to TPA is evident as the slope $\simeq2$,
indicating that $I_{UC}\propto I^{2}$.}

\par\end{centering}

\end{figure}

From eq. (\ref{eq:quadratic}), we expect a quadratic behavior of
$I_{UC}$ as we increase the peak power of the
excitation laser. This is experimentally confirmed in Fig. \ref{fig:Quadratic}
where the total fluorescence of the NQDs in a reference sample exhibits such quadratic behavior.

\begin{figure}
\begin{centering}
\includegraphics[width=0.7\columnwidth]{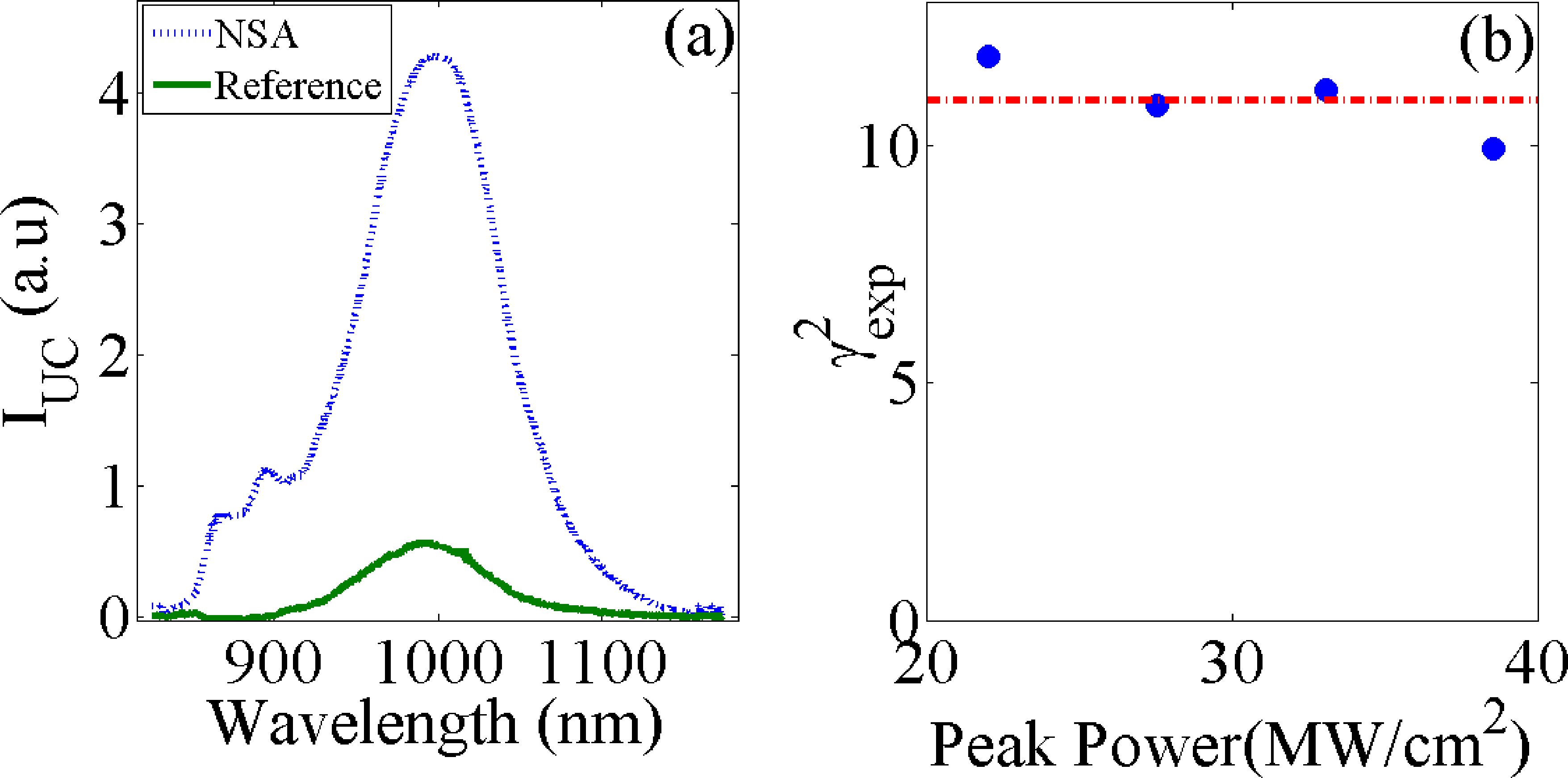}
\par\end{centering}

\caption{\label{fig:enhancement_and_equal}(a) Enhancement measurement with
an excitation at $1500nm$ in TM polarization. The dashed blue curve corresponds
to $I_{UC}$ from the investigated NSA sample while the green curve corresponds
to the reference sample. (b) $\gamma^2_{exp}$ as a function of the laser peak power showing that $\gamma^2_{exp}$ is power independent.}

\end{figure}

Fig. \ref{fig:enhancement_and_equal}a shows an example of an enhancement
of $I_{UC}$ in our NSA structures (in TM polarization). It is clear that the
induced fluorescence is enhanced due to the enhanced TPA. The square of the experimental enhancement
intensity factor ($\gamma^2_{exp}$) is extracted by integrating
the full spectra to get the total induced fluorescence and normalizing it to the emission from the reference sample.
In addition, we verified that $\gamma$  is power independent as expected, which is presented in Fig. \ref{fig:enhancement_and_equal}b.

\begin{figure}
\begin{centering}
\includegraphics[width=0.7\columnwidth]{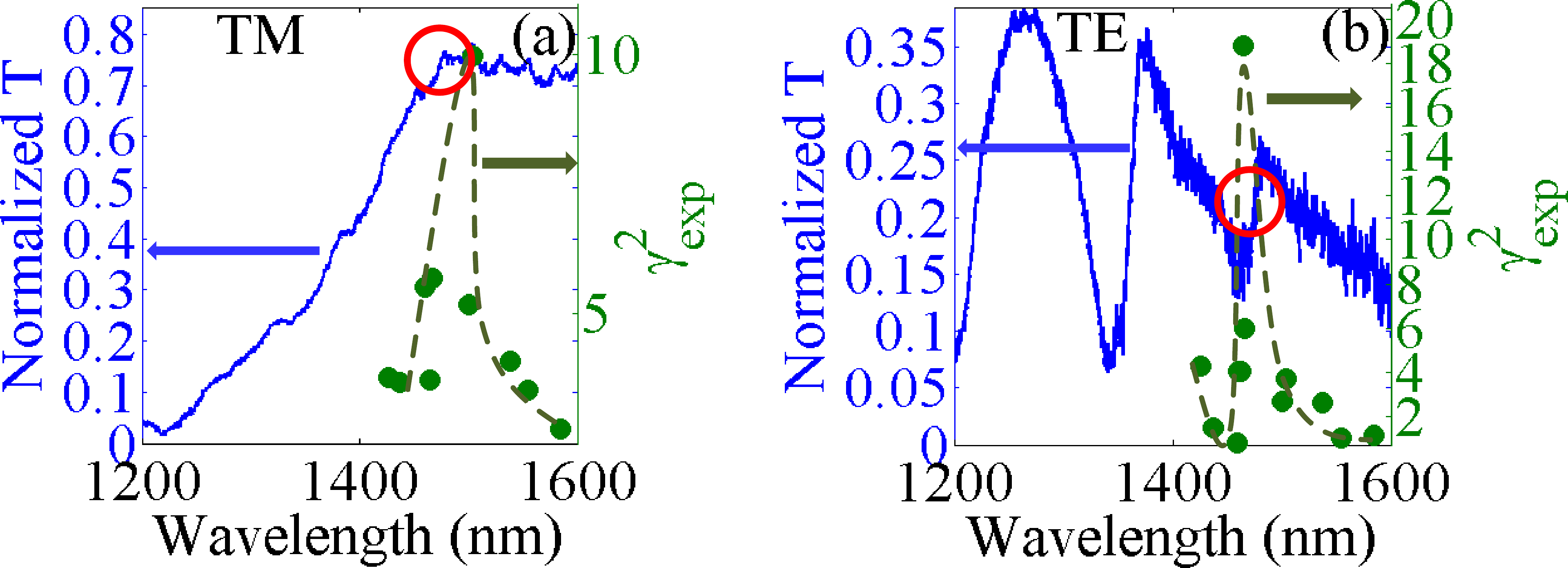}
\par\end{centering}

\caption{\label{fig:ST_enhance}The measured enhancement ($\gamma_{exp}^2$) for (a) TM polarization and (b) TE polarization is shown by the green dots as a function of the excitation wavelength (with a green dashed line as a guide to the eye). The transmission spectra from the respective NSA is shown by the blue curves. The red circles present the wavelengths used for the calculations in Fig. \ref{fig:ST_TE_dyn_diff}.}

\end{figure}

\begin{figure}
\begin{centering}
\includegraphics[width=0.7\columnwidth]{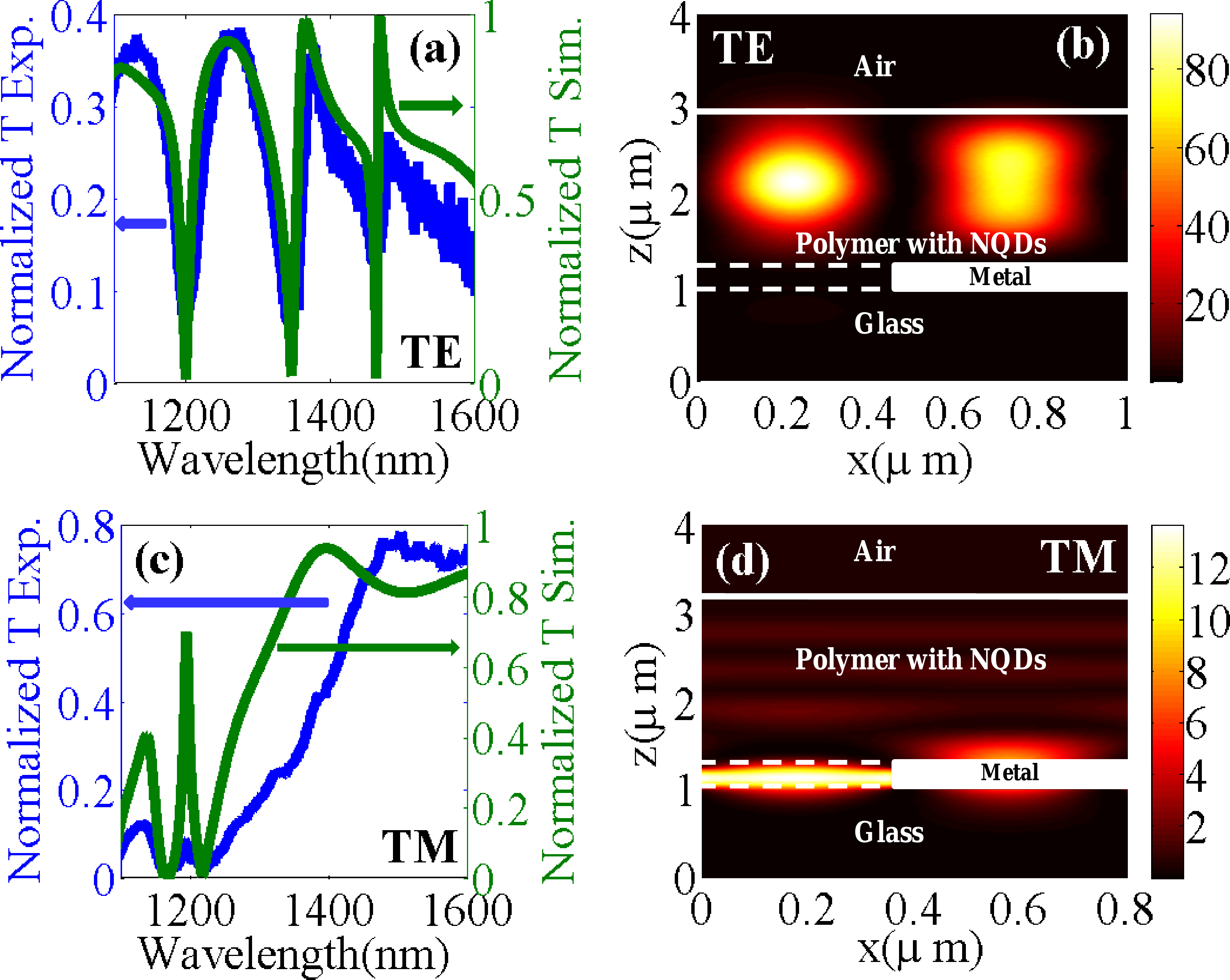}
\par\end{centering}

\caption{\label{fig:ST_TE_dyn_diff}
(a) and (c) - A calculated (green curves) and experimental (blue curves) transmission spectra for TE and TM polarizations, respectively. (b) and (d) - Calculated near-field intensities for the wavelengths indicated in Fig. \ref{fig:ST_enhance} by red circles for TE and TM polarizations respectively. The borders between the layers of the unit cell are indicated by the wight lines.}

\end{figure}

Next, we measured the dependence of $\gamma^2_{exp}$ on the excitation wavelength. This dependence is presented by the green dots in Fig. \ref{fig:ST_enhance}a for the TM excitation and in Fig. \ref{fig:ST_enhance}b for TE excitation. It is evident that in both TE and TM, the nonlinear enhancement is strongly wavelength dependent and exhibits a resonant behavior.
In order to understand the nature of this resonant behavior, we have measured the transmission spectra of the NSA structure in both polarizations. The corresponding transmission measurements in TM and TE are presented by the blue lines in Fig. \ref{fig:ST_enhance}a and b respectively. Both transmission spectra show clear EOT resonances. These peaks in transmission are the result of the standing EM resonances of the confined bragg modes of the structure (see Ref. \cite{Ilai} for a general analytical model that identifies and explains the origin of the EOT resonances in such structures). These Fabri-Perot like confined resonances are accompanied by a strong enhancement of the EM intensity inside the structure. This is verified by a numerical calculation of the optical response of our structures, based on a dynamical diffraction formalism \cite{Treacy}.  Fig. \ref{fig:ST_TE_dyn_diff}a,c presents a comparison of the measured transmission spectra to the calculated one for both TE and TM respectively. A good agreement is found between the calculation and the measurement, and the various peaks are identified as various orders of the standing wave resonances. In particular, we have calculated the EM intensity distributions in the structure around the first EOT resonance for both polarizations. The red circles in Fig \ref{fig:ST_enhance} point to the spectral position of the maximal calculated field enhancement for the TE and TM polarizations. It can be seen from Fig. \ref{fig:ST_enhance} that there is a good correspondence between the spectral location of the peak in $\gamma^2$ and the maximal field enhancement in both polarization. This is a strong indication that the mechanism behind the TPA enhancement is the local field enhancements at the EOT resonances. The calculated spatial distribution of the field intensities at those EOT resonances are plotted in Fig. \ref{fig:ST_TE_dyn_diff} b,d for TE and TM respectively, normalized to the EM field intensities without the NSA structure. It can be seen that indeed strong local field enhancements are obtained in our structures, reaching a maximal value of $\sim 90$ for TE and $\sim 13$ for TM.

To compare the  theoretical calculations to our experimental results, we first need to obtain the expected nonlinear enhancement factor from the near field calculations. For that, we assumed that the NQD are evenly dispersed in the polymer layer on the metallic grating. with this assumption, the spatially averaged value of the TPA enhancement factor
($\gamma_{calc}$) should be given by:

\begin{equation}
\gamma_{calc}(\lambda)=\frac{1}{\varepsilon_{PFCB}}\frac{\intop\limits _{unit-cell}I(r,\lambda)\,\mathbf{d}\bar{r}}{\intop\limits _{unit-cell}\,\mathbf{d}\bar{r}}
\label{gamma_calc}
\end{equation}
where $I(r)$ is the calculated field intensity enhancement in each point in the polymer layer in a unit cell of the NSA structure
and $\varepsilon_{PFCB}$ is the dielectric constant of the polymer $\varepsilon_{PFCB}$.

\begin{figure}
\begin{centering}
\includegraphics[width=0.4\columnwidth]{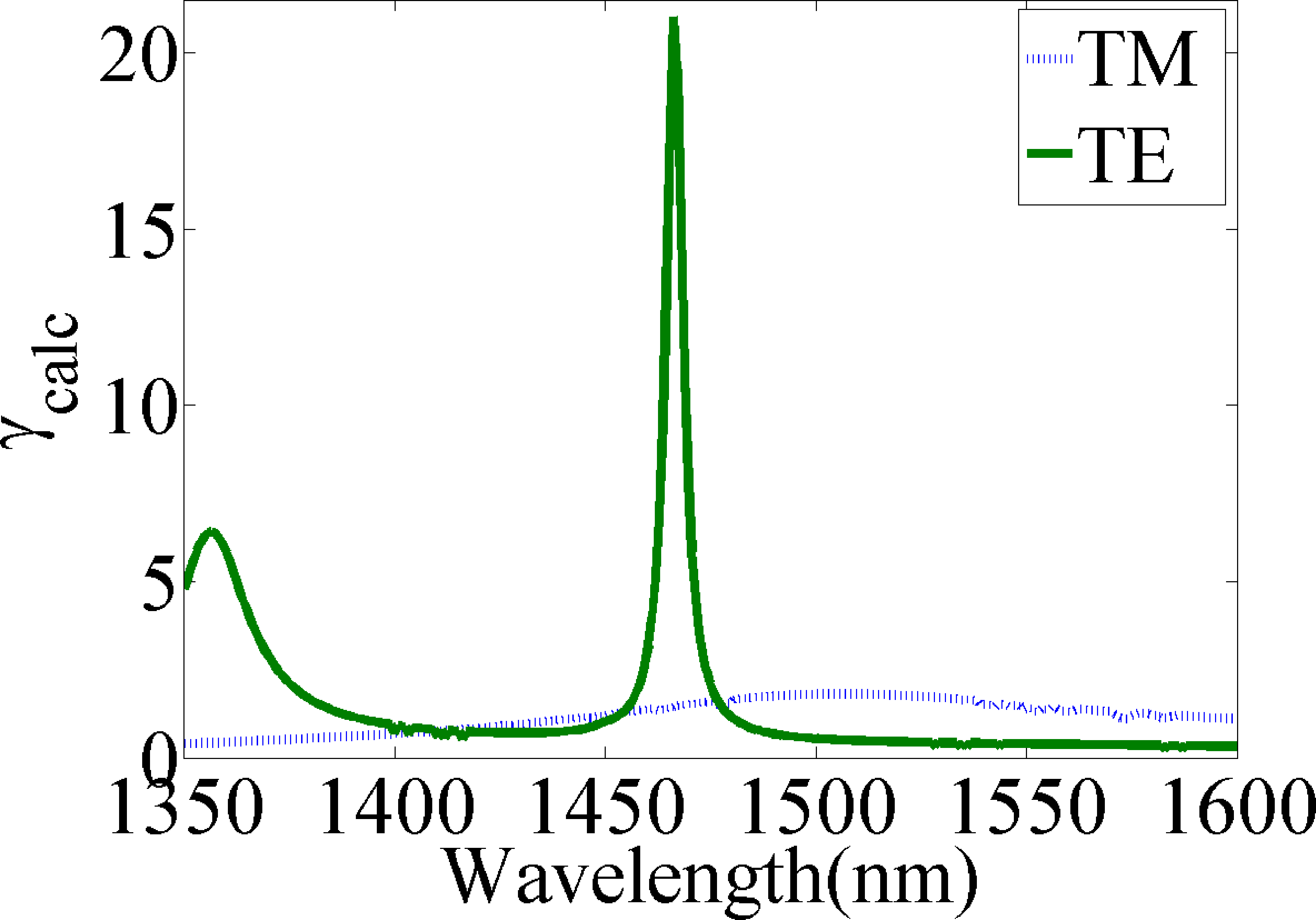}
\par\end{centering}

\caption{\label{fig:gamma_calc}$\gamma_{calc}$ as a function of the excitation wavelength.
The green curve corresponds to the calculated average enhancement
in TE polarization showing a very narrow peak ($\sim 3nm$ wide) at
$1466nm$. The dashed blue curve corresponds to TM polarization where the
magnitude is much lower than in TE polarization (peak value of $\sim 1.8$
at $1505nm$) with a width larger than $50nm$.}

\end{figure}

Fig. \ref{fig:gamma_calc} shows the dependence of $\gamma_{calc}$
on the excitation wavelength. The largest enhancement for TE
is at $1466nm$ with a maximal value of $\sim 21$ and
a typical width of $3nm$. For the TM polarization the enhancement is broader, peaking at $1505nm$. Next, we have to take into account the finite spectral width of the laser pulse ($\sim 20nm$), which excites the inhomogeneously broadened spectrum of the NQDs inside the resonant structure. This means that different NQDs having a TPA resonance at different wavelengths covered by the laser pulse spectrum experience different laser powers as well as different local field enhancements. To take this effect into account we calculate the expected averaged TPA enhancement by:
\begin{equation}
\gamma_{avg}=\frac{\intop\gamma_{calc}(\lambda)P(\lambda)\,\mathbf{d}\lambda}{\intop P(\lambda)\,\mathbf{d}\lambda},
\end{equation}
where ($P(\lambda)$) is the excitation laser pulse spectrum and ($\gamma_{calc}(\lambda)$) taken from Eq.~\ref{gamma_calc}.

\begin{figure}
\begin{centering}
\includegraphics[width=0.8\columnwidth]{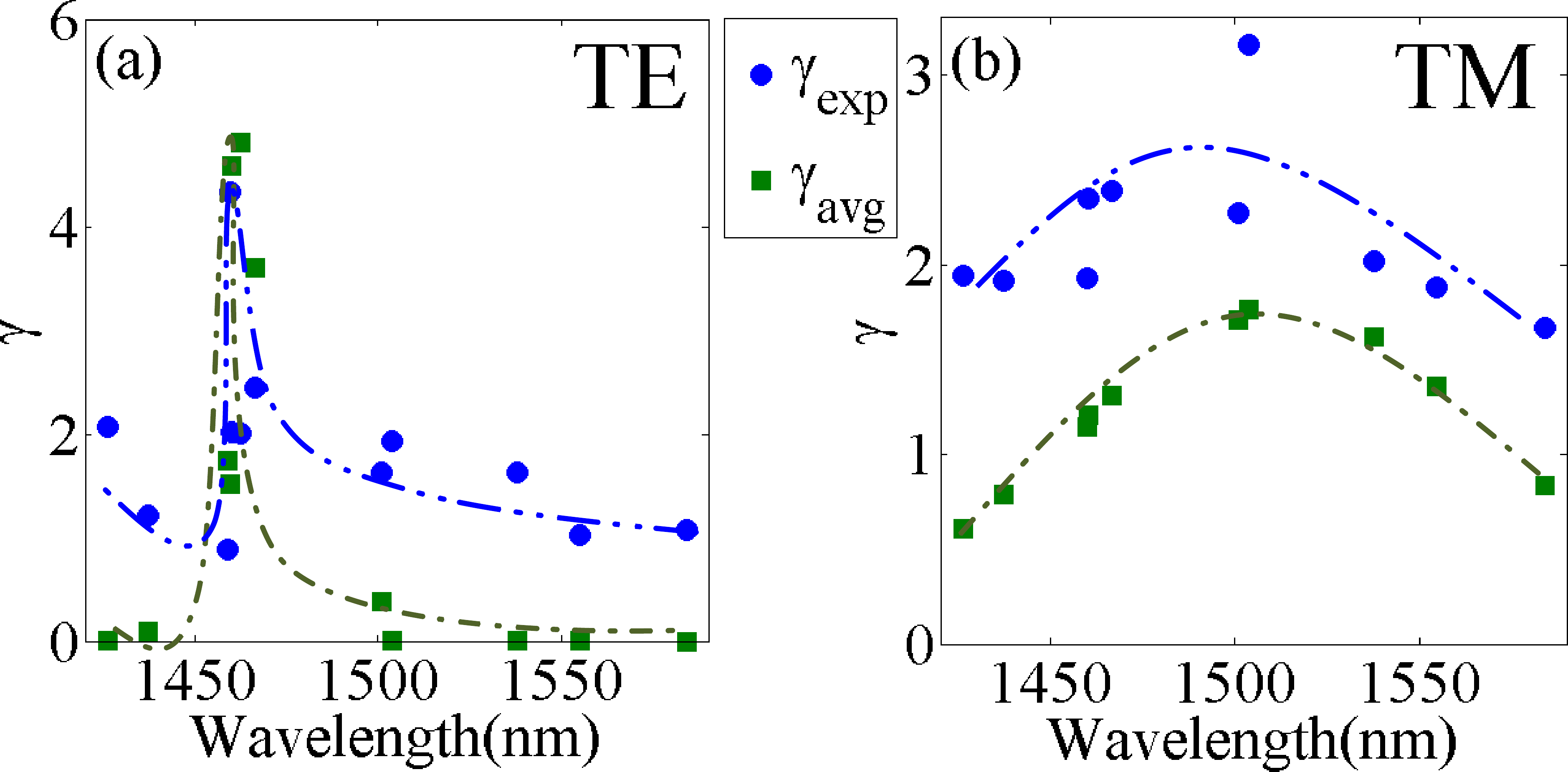}
\par\end{centering}

\caption{\label{fig:gamma_exp_avg}The experimental (blue dots) and the calculated
(green squares) enhancement factor as a function of the excitation
wavelength (the dashed lines are a guide to the eye) for (a) - TE polarization, (b) - TM polarization.}

\end{figure}

Fig. \ref{fig:gamma_exp_avg}a,b shows a comparison of the experimentally measured enhancement factor, $\gamma_{exp}$ to the calculated one, $\gamma_{avg}$, for TE and TM polarizations respectively. A very good agreement between the calculated values and the experimental values is obtained, confirming both the interpretation of the origin of the observed enhancement as well as validating the numerically calculated field intensities and the NQDs TPA enhancements in the NSA structure.

\section{Summary}

We have shown enhancement of two-photon absorption processes in nanocrystal quantum dots and of light upconversion from the IR to the NIR spectral regime using a hybrid optical device in which NIR emitting InAs quantum dots were embedded on top a a metallic nanoslit array. Our measurements and calculations show that the underlying mechanism behind the resonant enhancement of this nonlinear optical process is the strong local field enhancements inside the NSA structure which in turn result from the standing EM waves at the EOT resonances. A maximal TPA enhancement of more than 20 was inferred. We also identified, using our numerical model, the "hot spots" in the structure where maximal field enhancement is achieved: in TE polarization, the maximal field of the first EOT resonance occurs mainly at the dielectric waveguide layer on top of the metal grating, while for the first EOT resonance in TM, the maximal field occurs mostly inside the slit and very close to the metal-dielectric  interface. This can be used as a guide for designing and optimizing efficient nonlinear devices based on such NSA structures. For example, one can fill the slits and the top dielectric layer with different nonlinear media (such as two different types of NQDs), and control which one will be activated using different polarizations. We note that the maximal enhancement can be much improved by a better design, e.g., using less lossy metal in the IR spectral regime, and optimizing the metal thickness and the slit width. Our conclusion is that subwavelength metallic nanostructures are promising for a range of possible nonlinear optical devices based on nanocrystal quantum dots.

\bibliographystyle{unsrt}
\bibliography{bib}

\end{document}